\documentclass[aps,prl,twocolumn,showpacs,amsmath,amssymb,superscriptaddress]{revtex4-2}

\usepackage{graphicx}% Include figure files
\usepackage{dcolumn}% Align table columns on decimal point
\usepackage{bm}% bold math
\usepackage{color}

\usepackage{amsmath}
\usepackage{amssymb}
\usepackage{latexsym}
\usepackage{multirow}
\usepackage{xcolor}
\usepackage{hyperref}
\usepackage{comment}

\usepackage{textcomp}

\begin{document}

\title{Spontaneous oscillations and geometric cutoff in confined bacterial swarms}
\date{\today}
\author{Bing Miao}
\affiliation{Center of Materials Science and Optoelectronics Engineering, College of Materials Science and Opto-Electronic Technology, University of Chinese Academy of Sciences (UCAS), Beijing 100049, China}
\author{Lei-Han Tang}\email{tangleihan@westlake.edu.cn}
\affiliation{Department of Physics, Hong Kong Baptist University, Kowloon Tong, Hong Kong SAR, China}
\affiliation{Center for Interdisciplinary Studies, Westlake University, Hangzhou, China}

\begin{abstract}

Self-organized dynamic patterns in dense active matter are striking manifestations of non-equilibrium physics. A prominent example is the macroscopic elliptical motion observed in quasi-2D bacterial suspensions, which has lacked a physical explanation. Here, we examine a minimal linear response framework coupling bacterial swimming dynamics with fluid flow, treating long-range hydrodynamic interactions as a macroscopic communication channel. We demonstrate that microscopic swim motion, via Jeffery coupling, manifests as a ``phase-leading'' response to local shear flows. System-wide sustained oscillations, on the other hand, require both a critical bacterial density and strict geometric confinement. By analytically predicting the onset cell density and maximum film thickness, our model achieves excellent quantitative agreement with experiments, establishing a unified physical framework for self-organized periodic motion of elongated bodies in active fluids.
\end{abstract}

\maketitle

{\em Introduction} — 
Active fluids are fundamentally characterized by the continuous injection of energy at the scale of their individual constituents, driving them far from thermodynamic equilibrium~\cite{lauga,joanny13,rmp16}. In bacterial suspensions, for instance, this localized energy injection manifests through the persistent, run-and-tumble swimming dynamics of individual cells~\cite{yeomans,gompper,Saintillan18}. A central theme in the statistical mechanics of active matter is how these microscopic, non-equilibrium activities translate into macroscopic fluid behavior~\cite{ramaswamy,vicsek,Goldstein2012}. At sufficiently large spatial and temporal scales, the complex trajectories of individual agents typically coarse-grain: the persistent motion of bacteria randomizes over many tumbling events, renormalizing into effective passive properties such as enhanced translational diffusion or altered viscosity~\cite{sokolov-aranson09}. In such a scenario, the active fluid often loses its distinct non-equilibrium signature, phenomenologically mimicking a passive, equilibrium-like continuous medium~\cite{WKST}.

However, this macroscopic passive limit can be destabilized when individual cells communicate. A prominent example is the collective, elliptical motion of active bacteria observed above a critical cell density in a thin liquid film of 5 to 10 $\mu$m thickness~\cite{Wu17} (see Fig.~1). Even at high cell volume fractions of approximately $20\%$, individual bacteria maintain their characteristic run-and-tumble motion with an effective tumbling rate. While a 2D effective model can reproduce the synchronized trajectories through prescribed angular couplings~\cite{Wu17}, it leaves the physical origin of these interactions unresolved. 

\begin{figure}[htbp]
\label{fig:active_film}
\centering
  \includegraphics[width=1\linewidth]{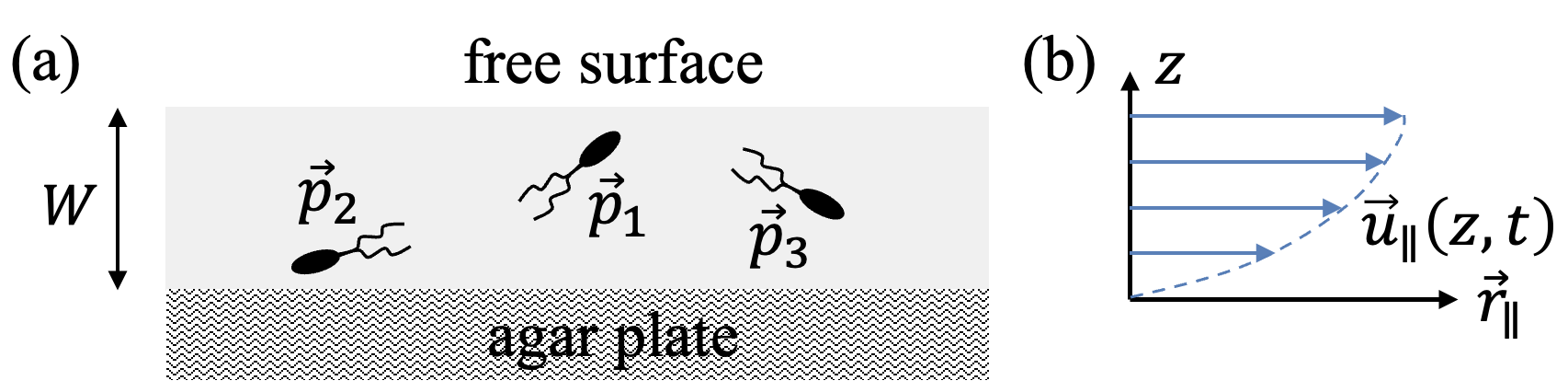}
\caption{\label{fig:active_film} 
(a) An active film of bacterial swimmers in a thin liquid layer atop an agar plate. Above a critical cell density, their coarse-grained velocity field exhibits coherent elliptical oscillations parallel to the plate, while individual cells maintain run-and-tumble dynamics, as reported in Ref.~\cite{Wu17}. (b) Proposed time-dependent, laminar flow profile that serves as a communication channel for alignment among swimmers.}
\end{figure}

In this Letter, we move beyond phenomenological descriptions to establish a first-principles hydrodynamic theory. Building on a recently formulated response-function framework~\cite{Wang-Tang}, we model the bacterial population as active agents that interact and communicate via the embedding fluid. In this view, macroscopic spontaneous oscillations emerge when the gain from cellular signal relay overcomes environmental dissipation—requiring both a collective ``phase-leading'' response and sufficient cell density. We demonstrate that these emergent oscillations do not require explicit behavioral programming; rather, they arise naturally as a rigorous consequence of long-range hydrodynamic interactions, governed by active stress generation and the Jeffery torque in a statistical sense. Furthermore, by explicitly incorporating the vertical boundary conditions of the experimental setup, our framework reveals that strict geometric confinement—specifically, the fluid film thickness—is not merely a secondary boundary effect, but the fundamental control parameter dictating the onset of instability.

{\em The model} — 
We begin by establishing the coupled dynamic equations for the bacteria and the fluid. The time evolution of the orientation-resolved single-particle distribution function $\rho(\vec{r},\vec{p},t)$, where the state of a bacterium is characterized by its position $\vec{r}$ and unit orientation vector $\vec{p}$, is governed by a Smoluchowski equation representing conservation of probability~\cite{shelley08-1}:
\begin{equation}
\label{smol}
\partial_t \rho(\vec{r},\vec{p},t)=-[\nabla\cdot \mathbf{J}_{\rm T}(\vec{r},\vec{p},t)+\nabla_{\vec{p}}\cdot \mathbf{J}_{\rm R}(\vec{r},\vec{p},t)].
\end{equation}
Here, $\mathbf{J}_{\rm T}$ and $\mathbf{J}_{\rm R}$ are the respective  translational and rotational fluxes, each consisting of a deterministic drift and a stochastic diffusion:
\begin{eqnarray}
\mathbf{J}_{\rm T}&=&\rho \mathbf{V}_{\rm T} = (v_{\rm s}\vec{p}+\vec{u})\rho -D_{\rm T}\nabla\rho, \nonumber\\
\mathbf{J}_{\rm R} &=& \rho \mathbf{V}_{\rm R} = \dot{\vec{p}}\rho - D_{\rm R} \nabla_{\vec{p}}\rho.\nonumber
\end{eqnarray}
In the translational flux, $v_{\rm s}$ represents the intrinsic swimming speed of the bacterium, while $\vec{u}(\vec{r})$ is the local fluid velocity field that advects the cells. In the rotational flux, the orientational drift of a single cell in a gradient fluid flow follows the Jeffery equation~\cite{jeffery,Saintillan18}: 
\begin{equation}
    \dot{\vec{p}} = \frac{1}{2}\boldsymbol{\Omega} \times \vec{p} + \beta \left[ \mathbf{E} \cdot \vec{p} - (\vec{p} \cdot \mathbf{E} \cdot \vec{p})\vec{p} \right].\nonumber
\end{equation}
Here $\mathbf{E} = \frac{1}{2}[\nabla\vec{u} + (\nabla\vec{u})^T]$ is the rate-of-strain tensor of the fluid, $\boldsymbol{\Omega} = \nabla \times \vec{u}$ is the fluid vorticity, and $\beta$ is a shape parameter of the swimmer, which tends to 1 in the slender limit (see SM~\cite{supplemental} for details). Finally, $D_{\rm T}$ and $D_{\rm R}$ are the translational and rotational diffusion coefficients, respectively; these characterize the stochastic components of bacterial motion and can be determined via single-bacterium trajectory experiments~\cite{gompper}.

The fluid flow field is governed by the Stokes equation:
\begin{equation}
\label{stok}
-\mu \nabla^2\vec{u}(\vec{r},t) = \nabla\cdot\mathbf{\Sigma}(\vec{r},t).
\end{equation}
Here, we have assumed a low Reynolds number regime where inertia is negligible, resulting in an instantaneous force-balance equation, with $\mu$ being the fluid viscosity and $\mathbf{\Sigma}(\vec{r},t)$ representing the active stress tensor exerted on the fluid by the swimming bacteria. The latter is defined as the orientational average of individual force dipoles, 
$$\mathbf{\Sigma}(\vec{r},t)=\sigma_0\int d^2\vec{p}\rho(\vec{r},\vec{p},t)\bigl[\vec{p}\vec{p}-\frac{1}{3}\mathbf{I}\bigr],$$ 
where $\sigma_0<0$ for pusher bacteria, and $d^2\vec{p}=\sin\theta d\theta d\phi$ is the solid angle. 

To align with the experimental setup illustrated in Fig.~\ref{fig:active_film}, we shall focus on laminar shear flow 
$$\vec{u}(z,t)=u_+(z,t)\hat{e}_++u_-(z,t)\hat{e}_-,$$ 
where $\hat{e}_{\pm}=(\hat{x}\pm i\hat{y})/\sqrt{2}$ are the basis vectors for circularly polarized fluid motion in the plane. By expressing the unit vector $\vec{p}$ in spherical coordinates $(\theta,\phi)$, Eqs. (\ref{smol}) and (\ref{stok}) respectively take their explicit forms:
\begin{eqnarray}
&&\bigl(\partial_t+v_{\rm s}\cos\theta\partial_z-D_{\rm T}\partial_z^2-D_{\rm R}\nabla^2_{\vec{p}}\bigr)\rho\nonumber\\
&=&\frac{3\beta\rho}{\sqrt{2}}(e^{i\phi}\partial_z u_+ + e^{-i\phi}\partial_z u_-)\sin\theta\cos\theta \nonumber\\
&&- \frac{\partial_\theta\rho}{2\sqrt{2}}(e^{i\phi}\partial_z u_+ + e^{-i\phi}\partial_z u_-)\left[1 + \beta(\cos^2\theta - \sin^2\theta)\right]\nonumber\\
&&- \frac{i(1+\beta)\partial_\phi\rho}{2\sqrt{2}}\frac{\cos\theta}{\sin\theta}(e^{i\phi}\partial_z u_+ - e^{-i\phi}\partial_z u_-),
\label{eq:rho_t}
\end{eqnarray}
\begin{equation}
\mu\partial_z u_\pm =-{\sigma_0\over\sqrt{2}}\int d^2\vec{p}\sin\theta\cos\theta e^{\mp i\phi}\rho(z,\theta,\phi,t).
\label{eq:stokes-2}
\end{equation}
Here $\nabla^2_{\vec{p}}={1\over\sin\theta}{\partial\over\partial\theta}\bigl(\sin\theta{\partial\over\partial\theta})+{1\over\sin^2\theta}{\partial^2\over\partial\phi^2}$. In the limit of low Reynolds number, we adopt the standard no-slip boundary condition, $u_\pm(z=0,t)=0$, at the bottom surface and a stress-free boundary condition, $\partial_z u_\pm(z,t)=0$ at the top surface ($z=W$). 

{\it Linear response in multipole expansion}  —  
Assuming the flow field $u_\pm(z,t)$ is weak, we treat the right-hand-side of Eq.~(\ref{eq:rho_t}) as perturbations to the static solution. As is typically done, we proceed by expanding the density distribution into spherical harmonics: 
\begin{equation}
\rho(z,\theta,\phi,t)=\sum_{l,m}a_{lm}(z,t)Y_l^m(\theta,\phi).
\label{eq:multipole}
\end{equation}
In this representation, the active stress driving the fluid in Eq.~(\ref{eq:stokes-2}) is governed exclusively by the off-diagonal nematic order parameter, $a_{2,\pm 1}$. Drawing an analogy to the signalling framework in Ref.~\cite{Wang-Tang}, we identify these specific harmonics as the macroscopic “communication channels” between the cells and the embedding fluid. In return, the resulting fluid flow $u_\pm(z,t)$ reorients the cells via the Jeffery torque in Eq. (\ref{eq:rho_t}), feeding the signal back into the orientational distribution of the swarm. Spontaneous oscillations emerge when the gain of this feedback cycle is positive.

We illustrate the scheme by first considering the simplified case of an isotropic and uniform density profile with $a_0 = c/\sqrt{4\pi}$, where $c$ is the mean cell density inside the liquid film. The spontaneous flow profile is modeled as $u_\pm(z,t)=\tilde{u} e^{-i\omega t/\tau}\sin(k z)$. Here, the frequency $\omega$ is made dimensionless by the tumbling timescale $\tau=1/D_{\rm R}$, and the wavenumber is set to $k=\pi/(2W)$, representing the lowest order mode compatible with the boundary conditions for $u_\pm(z,t)$.
Correspondingly, the density fluctuations in the $z$-direction are given by $a_{l,\pm 1}(z,t)=\tilde{a}_{l}e^{-i\omega t/\tau}\cos(kz)$ for even $l$ and $a_{l,\pm 1}(z,t)=\tilde{a}_{l}e^{-i\omega t/\tau}\sin(kz)$ for odd $l$. The resulting linearized dynamics in eigenmode space are governed by:
\begin{equation}
\label{modespace}
(-i\omega +\Lambda_l)\tilde{a}_l + g_l\tilde{a}_{l-1} - g_{l+1}\tilde{a}_{l+1}=B\delta_{l,2},\  l=1,2,\ldots
\end{equation}
with $\tilde a_0=0$. The Jeffery term $B\equiv -({3\over 20\pi})^{1/2}(k/D_{\rm R})\beta c\tilde{u}$ is coupled to the nematic tilt parameter $\tilde{a}_2$ via Eq. (\ref{eq:stokes-2}):
\begin{equation}
\label{W}
B = -\Bigl(\frac{\beta c\sigma_0}{5\mu D_{\rm R}}\Bigr)\tilde{a}_2.
\end{equation}
The diagonal and off-diagonal matrix elements are given by
\begin{eqnarray}
\Lambda_l=\Lambda_{\rm T}+l(l+1),\qquad
g_l=(-1)^l\text{Pe}_{\rm s} \sqrt{\frac{l^2-1}{4l^2-1}},\nonumber
\end{eqnarray}
with $\Lambda_{\rm T}=D_{\rm T}k^2/D_{\rm R}$. The ``active P\'eclet number''
$\text{Pe}_{\rm s}={\pi\over 2}v_{\rm s}/(D_{\rm R} W)$ is the ratio of bacterial tumbling interval against the channel crossing time~\cite{gompper,Saintillan18}.

To generate some analytical insight, we note that Eq. (\ref{modespace}) can be cast in the matrix form ${\bf L}{\bf\tilde{a}}={\bf B}$, where ${\bf\tilde{a}}=(\tilde{a}_1,\tilde{a}_2,\ldots)$, and ${\bf L}$ is a tri-diagonal, semi-infinite matrix whose inverse can be computed iteratively. Let $\psi_l$ be the determinant of the bottom-right submatrix of ${\bf L}$ from the diagonal element $L_{l,l}$ onward.
It satisfies $\psi_l=(-i\omega+\Lambda_l)\psi_{l+1}+g_{l+1}^2\psi_{l+2}$.
In terms of the ratio $x_l\equiv \psi_{l+1}/\psi_l$, we have
$x_l=(-i\omega + \Lambda_l + g_{l+1}^2 x_{l+1})^{-1}$,
which can be used to compute $x_l$ recursively from large $l$, where $\lim_{l\rightarrow\infty}x_l=0$. Following this procedure, we obtain
\begin{equation}
\tilde{a}_2 = ({\bf L}^{-1})_{22}B \equiv \chi(\omega, \Lambda_{\rm T}, \text{Pe}_{\rm s}) B,
\label{eq:def-chi}
\end{equation}
where the ``response function'' is given by,
\begin{equation}
\chi(\omega, \Lambda_{\rm T}, \text{Pe}_{\rm s})
={1\over x_2^{-1}+{1\over 5}(-i\omega+\Lambda_{\rm T}+2)^{-1}\text{Pe}_{\rm s}^2}.\label{response}
\end{equation}

{\em Results}  —  
Figure~\ref{fig:response}(a) displays the real and imaginary parts of $\chi\equiv \chi'+i\chi''$ against $\omega$ at $\Lambda_{\rm T}=1$ and two selected values of ${\rm Pe}_{\rm s}$. As expected, the imaginary part $\chi''(\omega)$ vanishes at zero frequency, as it is an odd function of $\omega$. For the P\'eclet number ${\rm Pe}_{\rm s}=5$, we observe $\chi''(\omega)>0$ for all $\omega>0$, indicating that fluctuations are dissipated and the system remains passive. However, increasing the P\'eclet number to ${\rm Pe}_{\rm s}=10$ drives $\chi''(\omega)$ to negative values at low-frequencies. Up to a cut-off frequency $\omega_0>0$, the bacterial population develops a phase-leading response to periodic shear flow.

\begin{figure}[htbp]
\centering
    \includegraphics[width=0.99\linewidth]{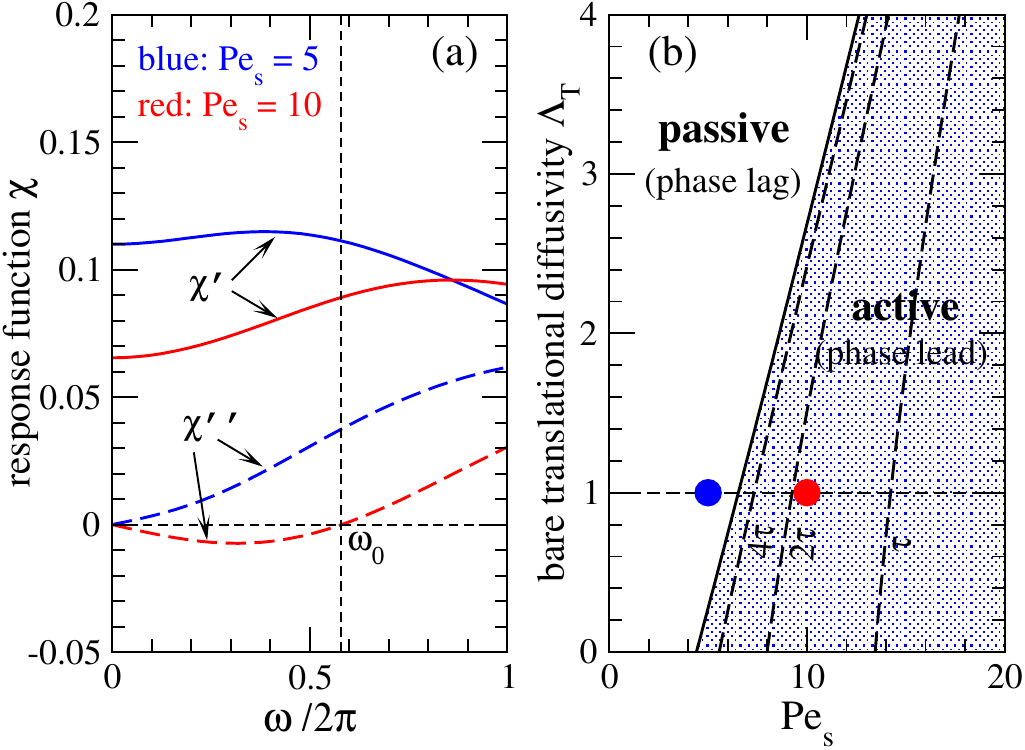}
\caption{\label{fig:response} Conditions for active response. (a) Frequency-resolved nematic tilt response function $\chi=\chi'+i\chi''$ to sinusoidal laminar shear flow at two different values of the P\'eclet number ${\rm Pe}_{\rm s}$. Here $\Lambda_{\rm T}=1$, and $\omega$ in units of $D_{\rm R}$. At ${\rm Pe}_{\rm s}=10$, $\chi''(\omega)<0$ for $0<\omega<\omega_0$, i.e., the cell population develops phase-leading response in this frequency range. 
(b) Regions of passive and active (shaded) response in the ${\rm Pe_s}$-$\Lambda_{\rm T}$ plane, separated by the solid line where the slope of $\chi''(\omega)$ vanishes at $\omega=0$. 
Contour lines (dashed) of $2\pi/\omega_0=\tau, 2\tau,\ldots$ are shown inside the active phase, with $\tau=1/D_{\rm R}$. }
\end{figure}

Figure~\ref{fig:response}(b) shows the ``phase diagram'' spanned by the two dimensionless parameters ${\rm Pe}_{\rm s}$ and $\Lambda_{\rm T}$.
The solid line separates the region with $\omega_0>0$ (shaded) from the passive phase with $\chi''(\omega)>0$ for all $\omega>0$.
The two examples in Fig.~\ref{fig:response}(a) lie on either side of this phase boundary, shown by the blue (${\rm Pe}_{\rm s}=5$) and red (${\rm Pe}_{\rm s}=10$) dots. Active response of swarming bacteria to laminar shear flow occurs only when ${\rm Pe}_{\rm s}$ is beyond a threshold value ${\rm Pe}_{\rm s}^{\rm c}(\Lambda_{\rm T})$ (solid line in Fig. ~\ref{fig:response}(b)). At a given ${\rm Pe}_{\rm s}>{\rm Pe}_{\rm s}^{\rm c}(0)=4.406\ldots$, the cutoff frequency $\omega_0$ for phase-leading response decreases with increasing $\Lambda_{\rm T}$, eventually vanishes at the phase boundary.

Spontaneous motion of the bacterial film further requires nontrivial solutions to Eq. (\ref{W}). 
Together with Eq. (\ref{eq:def-chi}), we obtain the minimal cell density $c_{0}$ at the onset of oscillations,
\begin{equation}
    c_0 = 
    -\Bigl(\frac{5\mu D_{\rm R}}{\beta\sigma_0}\Bigr)[\chi'(\omega_0,\Lambda_{\rm T},{\rm Pe}_{\rm s})]^{-1}.
    \label{c_0}
\end{equation}
Beyond this density, the torque generated by the cell population is sufficient to overcome viscous forces within the fluid, leading to the elliptical flow as observed in the experiment.

Experimentally, from the balance of the drag force $F_{\rm drag}\sim \mu v_{\rm s}$ with the propulsion strength $\sigma_0$, one may conclude that the ratio $\mu v_{\rm s}/\sigma_0$ is a constant, though there could be small variations depending on the bacterial shape and flagellar positions, and other factors. Typically it is expected to fall in the range $0.025 \sim 0.05\ \mu{\rm m}^{-2}$~\cite{Drescher2011,gompper}. In Ref.~\cite{Wu17}, it was reported $v_{\rm s}\simeq 34 \ \mu{\rm m/s}$. Furthermore, their trajectory analysis yielded a time constant $\tau_1=2.3$ sec for turns of $\Delta\phi>3\pi/4$, which translates to a rotational diffusivity $D_{\rm R}=2/\tau_1\simeq 0.87\ {\rm s}^{-1}$ when the run direction is randomized upon each tumbling event. 

Figure~\ref{fig:density}(a) shows the minimal cell density $c_0$ against the active P\'eclet number $\text{Pe}_\text{s}$, calculated using Eq.~(10) with the aforementioned experimental parameters. The corresponding onset frequency $\omega_0/2\pi$ is displayed in Fig.~\ref{fig:density}(b). Here, the shape parameter $\beta$ is set to 1 to model \textit{B.~subtilis}. On the other hand, the translational diffusivity $D_\text{T}$ is generally assumed to be small due to the cells' ballistic motion at short timescales~\cite{rmp16,gompper}. For a dimensionless translational diffusivity in the range $0 \le \Lambda_\text{T} \le 1$ (indicated by the shaded regions), the predicted critical density $c_0 \simeq 0.05~\mu\text{m}^{-3}$ closely matches the experimental findings reported in Ref.~\cite{Wu17} (solid red line). On the other hand, the onset frequency exhibits a strong dependence on the film thickness $W$. In the experiment, the 
film thickness ranges from $5~\mu\text{m}$ to $10~\mu\text{m}$, as highlighted by the vertical blue dashed lines. At a given $\Lambda_{\rm T}$, the oscillation frequency decreases as the film width $W$ increases (corresponding to a decrease in $\text{Pe}_\text{s}$), ultimately vanishing at a maximum film thickness slightly above $10~\mu\text{m}$.

\begin{figure}[htbp]
\centering
\includegraphics[width=\linewidth]{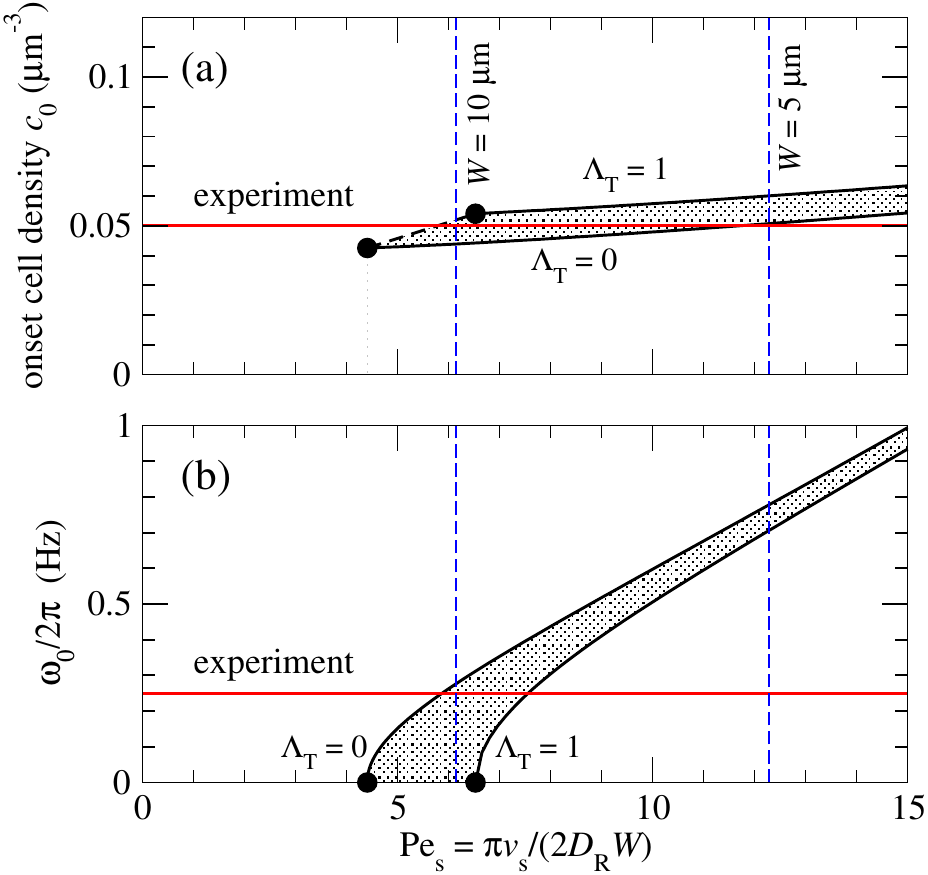}
\caption{\label{fig:density} 
Comparison of theoretical predictions ($\beta = 1$) with experimental observations from Ref.~\cite{Wu17} (solid red lines) for (a) the onset cell density $c_0$ and (b) the oscillation frequency $\omega_0/2\pi$. The shaded regions represent the theoretical predictions bounded by the dimensionless translational diffusivity range $0 \le \Lambda_\text{T} \le 1$. The vertical blue dashed lines indicate the range of active film thicknesses ($W = 5~\mu\text{m}$ to $10~\mu\text{m}$) where spontaneous motion was experimentally observed. Notably, the model predicts that spontaneous oscillations cease when the film exceeds a maximum thickness slightly above $10~\mu\text{m}$. }
\end{figure}

{\em Conclusions} — By coupling the dynamics of swimming bacteria with fluid flow through a macroscopic linear response framework, we have elucidated the physical mechanism underlying self-sustained elliptical motion in quasi-2D active suspensions. Our analysis demonstrates that, with a judicious combination of bacterial thrust $v_{\rm s}$ and the Jeffery rotation, microscopic active energy is injected to the fluid that manifests as a macroscopic ``phase-leading'' response to local shear flows. With the fluid flow as the ``organizer'', the bacterial nematic stress is channeled coherently to drive the dissipative fluid. 

For this microscopic energy injection to trigger a macroscopic instability, three fundamental conditions must align: the cells must respond with the correct phase to amplify the flow, the population density must exceed a critical threshold to overcome environmental viscous damping, and the system must be strictly confined transverse to the flow. The latter aspect is particularly interesting: if the channel width exceeds the typical length of a single bacterial run (effectively lowering the active P\'eclet number ${\rm Pe_s}$), cells lose their orientational memory before traversing the full length of the film, and thus significantly weaken their coherent macroscopic response. One may also note that, multiple tumbles increase the translational diffusivity $D_{\rm T}$ as well, thereby smearing out the polarization profile $a_{l,\pm 1}(z,t)$ and pushing up the instability threshold ${\rm Pe_s^c(\Lambda_{\rm T})}$. Our work thus elevates the fluid film thickness from a mere boundary condition to the primary control parameter of spontaneous flow. Similar confinement-induced transitions—where boundaries suppress active fluctuations to stabilize coherent, large-scale flows—have also been established in the broader context of active nematics~\cite{Yeomans2017}.

By framing long-range hydrodynamic interactions as a macroscopic communication channel, our model bridges active kinetic theory with the physics of collective synchronization~\cite{Wang-Tang}. The excellent quantitative agreement with experimental onset conditions and frequencies strongly validates this first-principles approach. While our linear stability analysis successfully captures the onset of this collective behavior, it naturally leaves the saturated state open. Extending this theoretical framework into the nonlinear regime will eventually clarify the amplitude saturation and long-term topological stability of these emergent orbits, paving the way for a complete thermodynamic description of self-organized periodic motion in confined active matter.

{\em Acknowledgments} — L.T. wishes to thank Dr. Yilin Wu for many useful discussions, and Dr. Francois Gu who participated in the early part of the project during a summer visit to HKBU. B.M. thanks the Center for Interdisciplinary Studies at Westlake University for hospitality, where the collaboration was initiated. The work is supported by the Research Grants Council of the HKSAR under Grant No. 12304020, and by the Ministry of Education of China under grant JYB2025XDXM502. B.M. acknowledges support from the National Natural Science
Foundation of China (NSFC) (Grant No. 12575045).

\clearpage % Forces the supplement to start on a new page
\onecolumngrid % Switches to single-column format

% --- Formatting the Supplementary Title ---
\begin{center}
    \textbf{\large Supplemental Material for: Spontaneous oscillations and geometric cutoff in confined bacterial swarms}\\[.2cm]
    Bing Miao and Lei-Han Tang
\end{center}
\vspace{1cm}

% --- Resetting Counters and Adding 'S' Prefix ---
\setcounter{equation}{0}
\setcounter{figure}{0}
\setcounter{table}{0}
\setcounter{page}{1}
\makeatletter
\renewcommand{\theequation}{S\arabic{equation}}
\renewcommand{\thefigure}{S\arabic{figure}}
\renewcommand{\thetable}{S\arabic{table}}
\makeatother

\title{Supplementary Material for: ``Spontaneous oscillations and geometric cutoff in confined bacterial swarms''}

\maketitle

The general form of the Jeffery equation, which describes the evolution of the orientation vector \(\vec{p}\) for an elongated object in a three-dimensional fluid flow field \(\vec{u}(\vec{r}, t)\), is given by~[6,15]:
\begin{equation}
    \dot{\vec{p}} = \frac{1}{2}\boldsymbol{\Omega} \times \vec{p} + \beta \left[ \mathbf{E} \cdot \vec{p} - (\vec{p} \cdot \mathbf{E} \cdot \vec{p})\vec{p} \right],
    \label{eq:jeffery}
\end{equation}
where \(\mathbf{E} = \frac{1}{2}[\nabla\vec{u} + (\nabla\vec{u})^T]\) is the rate-of-strain tensor of the fluid, \(\boldsymbol{\Omega} = \nabla \times \vec{u}\) is the fluid vorticity, and \(\beta\) is Bretherton's constant. For a spheroidal object with aspect ratio \(\alpha\), the shape parameter is defined as \(\beta = (\alpha^2 - 1)/(\alpha^2 + 1)\), which approaches \(1\) in the slender limit (\(\alpha \to \infty\)).

In the main text, we consider a laminar shear flow parallel to the film surface, defined as:
\begin{equation}
    \vec{u}(z, t) = u_x(z, t)\hat{x} + u_y(z, t)\hat{y}.
    \label{eq:flow}
\end{equation}
For this specific flow profile, the velocity depends only on the $z$-coordinate, meaning the only non-zero components of the velocity gradient tensor are $\partial_z u_x$ and $\partial_z u_y$. Consequently, the rate-of-strain tensor $\mathbf{E}$ and the vorticity vector $\boldsymbol{\Omega}$ take the explicit forms:
\begin{equation}
    \mathbf{E} = \frac{1}{2} \begin{pmatrix} 
    0 & 0 & \partial_z u_x \\ 
    0 & 0 & \partial_z u_y \\ 
    \partial_z u_x & \partial_z u_y & 0 
    \end{pmatrix}, \quad 
    \boldsymbol{\Omega} = \begin{pmatrix} 
    -\partial_z u_y \\ 
    \partial_z u_x \\ 
    0 
    \end{pmatrix}.
    \label{eq:tensors}
\end{equation}

Expressing the orientation vector in angular coordinates, $\vec{p} = (\sin\theta\cos\phi, \sin\theta\sin\phi, \cos\theta)$, it follows that
\begin{equation}
    \frac{1}{2}\boldsymbol{\Omega} \times \vec{p} = \frac{1}{2} \begin{pmatrix} \partial_z u_x \cos\theta \\ \partial_z u_y \cos\theta \\ -S\sin\theta \end{pmatrix}, \quad
    \mathbf{E} \cdot \vec{p} = \frac{1}{2} \begin{pmatrix} \partial_z u_x \cos\theta \\ \partial_z u_y \cos\theta \\ S\sin\theta \end{pmatrix}.
    \label{eq:flow-tensor-explicit}
\end{equation}
Here, we have introduced the auxiliary scalar variable $S$, defined as:
\begin{equation}
    S = \cos\phi \ \partial_z u_x + \sin\phi \ \partial_z u_y = \frac{1}{\sqrt{2}}(e^{i\phi}\partial_z u_+ + e^{-i\phi}\partial_z u_-),
\end{equation}
with $u_\pm = (u_x \mp i u_y)/\sqrt{2}$. 

Since $\vec{p}$ is a unit vector, the orientation rate $\dot{\vec{p}}$ can be decomposed into angular velocity components along the local spherical basis vectors $\hat{\theta} = (\cos\theta\cos\phi, \cos\theta\sin\phi, -\sin\theta)$ and $\hat{\phi} = (-\sin\phi, \cos\phi, 0)$. 
With the help of Eq. (\ref{eq:flow-tensor-explicit}), the resulting angular velocity components are given by:
\begin{eqnarray}
    \dot{p}_\theta &=& \dot{\vec{p}} \cdot \hat{\theta}=\frac{1}{2}S \left[ 1 + \beta(\cos^2\theta - \sin^2\theta) \right],\\
    \dot{p}_\phi &=& \dot{\vec{p}} \cdot \hat{\phi} = \frac{1 + \beta}{2}V \cos\theta.
\end{eqnarray}
Here
\begin{equation}
    V = - \sin\phi\ \partial_z u_x + \cos\phi\ \partial_z u_y = \frac{i}{\sqrt{2}}(e^{i\phi}\partial_z u_+ - e^{-i\phi}\partial_z u_-) \equiv \partial_\phi S .
\end{equation}

For an arbitrary shape parameter $\beta$, the rotational drift term on the right-hand side (RHS) of Eq. (3) in the main text is given by the negative divergence of the rotational flux:
\begin{equation}
    \text{RHS} = -\nabla_{\vec{p}} \cdot (\rho \dot{\vec{p}}) = -\frac{1}{\sin\theta}\partial_\theta(\rho \sin\theta \dot{p}_\theta) - \frac{1}{\sin\theta}\partial_\phi(\rho \dot{p}_\phi).
    \label{eq:RHS}
\end{equation}

Substituting the expressions for $\dot{p}_\theta$ and $\dot{p}_\phi$ into Eq. (\ref{eq:RHS}) and noting that $\partial_\theta S =0$ and $\partial_\phi V= - S$, we obtain,
\begin{equation}
    \text{RHS} = -\frac{S}{2 \sin\theta} \Bigl[\bigl(1 + \beta(\cos^2\theta - \sin^2\theta)\bigr)\sin\theta\partial_\theta\rho  + \bigl(1 + \beta(\cos^2\theta - 5 \sin^2\theta)\bigr)\rho \cos\theta\Bigr] - \frac{1 + \beta}{2 \sin\theta}\cos\theta\bigr(V\partial_\phi\rho - \rho S\bigr).
\end{equation}
By grouping the terms proportional to $\rho S$, the expression simplifies significantly:
\begin{equation}
    -\frac{\rho S \cos\theta}{2 \sin\theta} \bigl[ 1 + \beta(\cos^2\theta - 5 \sin^2\theta) - (1 + \beta) \bigr] = 3\beta\rho S \sin\theta \cos\theta.
\end{equation}

Consequently, for a general shape parameter $\beta$, the governing equation for the bacterial orientation distribution $\rho$ under gradient fluid flow takes the form:
\begin{equation}
\begin{aligned}
    (\partial_t + v_{\rm s} \cos\theta\partial_z - D_{\rm T}\partial_z^2 - D_{\rm R}\nabla_{\vec{p}}^2)\rho &= \frac{3\beta\rho}{\sqrt{2}}(e^{i\phi}\partial_z u_+ + e^{-i\phi}\partial_z u_-)\sin\theta\cos\theta \\
    &\quad - \frac{\partial_\theta\rho}{2\sqrt{2}}(e^{i\phi}\partial_z u_+ + e^{-i\phi}\partial_z u_-)\left[1 + \beta(\cos^2\theta - \sin^2\theta)\right] \\
    &\quad - \frac{i(1+\beta)\partial_\phi\rho}{2\sqrt{2}}\frac{\cos\theta}{\sin\theta}(e^{i\phi}\partial_z u_+ - e^{-i\phi}\partial_z u_-).
\end{aligned}
\end{equation}

\end{document}